\documentstyle[11pt,twoside,colloqOHP,epsfig]{article}
\begin{document}
\title{Planets of young stars: the TLS radial velocity survey}
\author{M. Esposito$^{1,2}$, E. Guenther$^1$, 
A.P. Hatzes$^1$, M. Hartmann$^1$}
\affil{$^1$Th\"uringer Landessternwarte Tautenburg, 
Sternwarte 5, 07778 Tautenburg, Germany} 
\affil{$^2$Dipartimento di Fisica ``E.R. Caianiello'', 
Universit\`a  di Salerno, via S. Allende, 84081 Baronissi 
(Salerno), Italy} 
\begin{abstract}
We report on the search for planets orbiting  46 nearby young stars 
performed at the State Observatory of Turingia (TLS) by means of a 
radial velocity survey. The aim of this program is to test the theories
 of formation/evolution of planetary systems. For 19(8) stars we can 
exclude planets with $M \sin i$ $\geq$ 1 $M_J$ (5 $M_J$) and 
$P$ $\leq$ 10 days; we find 1 short period binary and 5 stars with 
long period RV-trend. One good young exo-planet candidate is 
presented. 
\end{abstract}
\section{Introduction}

In the last ten years, thanks to high precision radial velocity (RV)
 measurements, more than 100 extrasolar planets have been found.
 The most important observational campaigns up to now have 
focused on old solar type stars because such stars present  all the
 characteristics (small $v\sin i$, high number of spectral lines, low 
activity) to exploit the best potentialities of the RV technique. 

The orbits of the extra-solar planets are strikingly different from 
those of our solar system: some of them have massive planets of 
extremely short orbital periods (so called ``hot Jupiters'' or 
``Pegasides''), while others exhibit very high eccentricities.  
These results have given a new strong impulse to the theoretical 
efforts to explain the formation and evolution of planetary systems. 
However, even the most fundamental questions are still open. How 
common are planetary systems?
How do planets form? By core accretion or by gravitational instability 
in the disk. Where do they form? Close-in Jupiter mass planets can 
form in situ, or do they have to form at a certain distance and then 
migrate inward? New insights should come from the determination of 
the orbital parameters for planets in early evolutionary phases. 
According to some scenarios the frequency of planets was initially 
much higher than is observed in old stars, as a substantial fraction 
of the planets might have been either ejected from the system, or 
have been engulfed in the host stars. Since capture of planets is 
highly unlikely, the frequency of planets of young stars ought to be 
higher than that of old stars. The aim of this survey is to find out 
by how much. Another issue we could address concerns the 
evolution of the orbital parameters. In particular it would be 
important to know whether close-in planets have  round orbits 
when they form or  eccentric orbits, which then get circularised by 
tidal interaction with the host star.

There is an additional reasons for searching for planets of young 
stars. As pointed out by Sudarsky, Burrows, \& Hubeny (2003) even 
old exo-planets orbiting a solar-like star at 0.1 AU would have a 
temperature of up to 900~K, because they are heated by the star. 
A massive, isolated planet with an age between $10^7$ 
to $10^8$ yrs would also have a temperature of more than 800~K 
but this time because it is still contracting. For close-in  young 
planets both effects would add up resulting in objects of spectral 
type early L or even late M, which would be only 5 to 7 mag fainter 
than the host star. The direct detection of these objects, especially 
in the infrared light, would be possible by means of interferometric 
observations as well as tracking down spectral signatures, giving 
access to fundamental parameters like temperature, radius and 
true masses.

The presence of spots and/or plages in the photosphere of active 
stars  changes the profile of  spectral lines causing fictitious 
variations in the RV measurements (the so-called jitter). Not only 
this source of noise can mask the periodical variations which are 
the characteristic signature of the presence of a planet, but what 
makes it even worse is the fact that the presence of spots 
combined with the stellar rotation can itself introduce spurious 
periodicities in the radial velocity signal.This explains why young 
stars, which are supposed to be active, have so far been excluded 
from the major RV surveys for planets detection. However Paulson 
et al. (2002) monitored 82 stars in the Hyades (age $\sim$ 700 Myr) 
finding a significant correlation between simultaneous RV and
$R'_ {HK}$ for only 5 stars. Indeed exactly to what extent the stellar 
activity can hinder the detection of exoplanets in RV measurement 
has not yet clearly assessed.

Weighing pros and cons we finally  decided to undertake a 
RV survey of young stars.

\section{The TLS radial velocity survey}

Observations began in 2001. We monitored a sample of 46 young, 
nearby dwarf  stars of late spectral type and took 1500 spectra 
of these up to now.

\subsection{The instrumental setup}
The observations have been carried out at the State Observatory 
of Turingia (TLS) using the 2m  ``Alfred Jensch'' telescope 
which is equipped with a Coud\'e echelle spectrograph.
We used the visual grism which cover the spectral region from 
4660 to 7410 $\AA$ in 44 orders. With a slit width of 1.2 arcsec 
a resolution of $R = 67000$ is achieved.
An iodine cell placed in front of the slit generates a very dense
 system of absorption lines superimposed onto the stellar spectrum, 
which provides a highly precise wavelength scale  and at same 
time allows to measure the
 PSF in situ over the spectrum. Radial velocities 
(RV's) are measured by means of  a software package called RADIAL 
developed at the University of Texas and McDonald Observatory, 
based on the methods described in Marcy \& Butler(1992) 
and Valenti, Butler \& Marcy(1995).

\begin{figure}[th]
\begin{center}
\epsfig{file=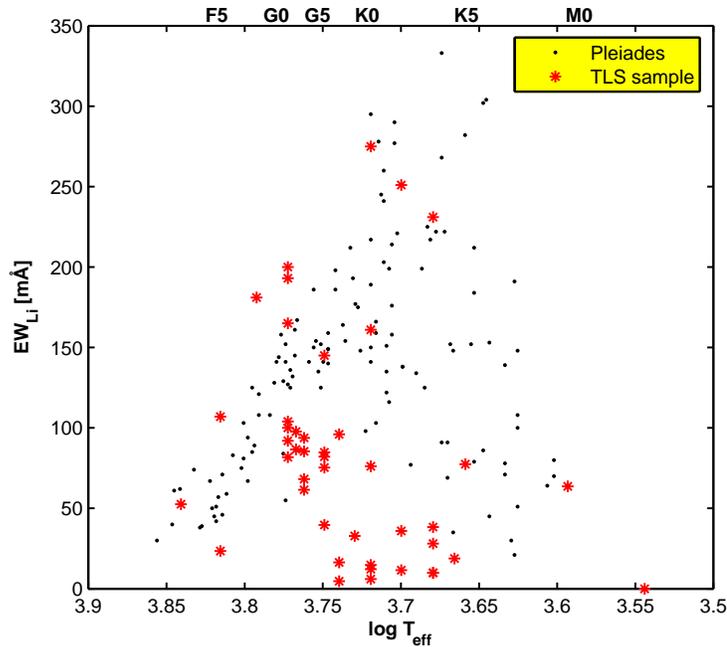,width=11cm}
\caption{Plot of the LiI$\lambda$6708 vs logT$_{\tiny{eff}}$ for 
the TLS sample and, as a comparison, for the Pleiades. Most of 
the star surveyed have ages comparable to the Pleiades 
($\sim100$ Myr); the clump with  $3.65 < logT_{\tiny{eff}} < 3.75$ 
and EW $< 50m\AA$ is probably made of  
$300 \div 500$ Myr old stars (see text).} 
\end{center}
\end{figure}

\subsection{Characterisation of the sample}

The identification of bona-fide young post-T-Tauri stars near the Sun 
still is an open astrophysical issue (Jensen 2001). Recently many 
nearby associations of young stars have been recognised: 
TWA, $\beta$ Pic, Tucana-Horologium, $\eta$ Cha, AB Dor 
(Zuckermann \& Song 2004). For such coeval stellar groups statistics 
considerations help to  achieve reliable estimations of the common 
age. Unfortunately all those associations are located in the Southern 
hemisphere.Thus in selecting the sample of stars to survey we did 
the Hobson's choice and looked out for young 'field' stars.

One of the best indicators of young age, especially for G and K stars, 
is the presence of the lithium LiI$\lambda$6708 absorption line. 
The convective envelope of these stars brings the lithium in contact 
with the stellar core where the temperature is high enough to cause 
its burning. As a consequence the lithium in the 
stellar atmospheres is progressively 
depleted. The lithium abundances can thus be used as an age estimator.
Fig.1 shows the LiI$\lambda$6708 equivalent width (EW) as a function 
of the $T_{\tiny{eff}}$ for the stars in our sample and, as a 
comparison, for the Pleiades (age $\sim100$ Myr).
As can be seen for the stars of Pleiades, the LiI$\lambda$6708 
EW scatters significantly for the same $T_{\tiny{eff}}$. Thus it is not 
possible  determine the age for every single star in our sample precisely. 
Rather, in a schematic way, we can subdivide our sample in two groups, 
one having ages comparable to the Pleiades and the other consisting of 
older stars. In fact, many stars in the latter group have been recognised 
as members of the Ursa Major association which is 300 Myr old 
(Soderblom \& Major 1993)  (500 Myr according to King \& Schuler 2005).

Based on the Hipparcos parallaxes all our stars are 
at a distance of less than 50 pc  
from the Sun, and 36 of them are closer than 30 pc. Accordingly, they 
are relatively bright, ranging from the fifth to the ninth visual magnitude.

\section{Analysis of the data}

\subsection{Internal errors}
Fig.2 (left panel) shows the average internal errors $\sigma_{int}$ 
in the RV's as a function of the projected stellar rotational velocity 
$v\sin i$. As expected for rapidly rotating stars, the spectral lines 
broadening makes it difficult to achieve high precision RV values.
 However, up to a $v\sin i \leq 10 $ km/s, we routinely get 
internal errors of 10 to 15 m/s . 

\begin{figure}[th]
\begin{center}
\plottwo{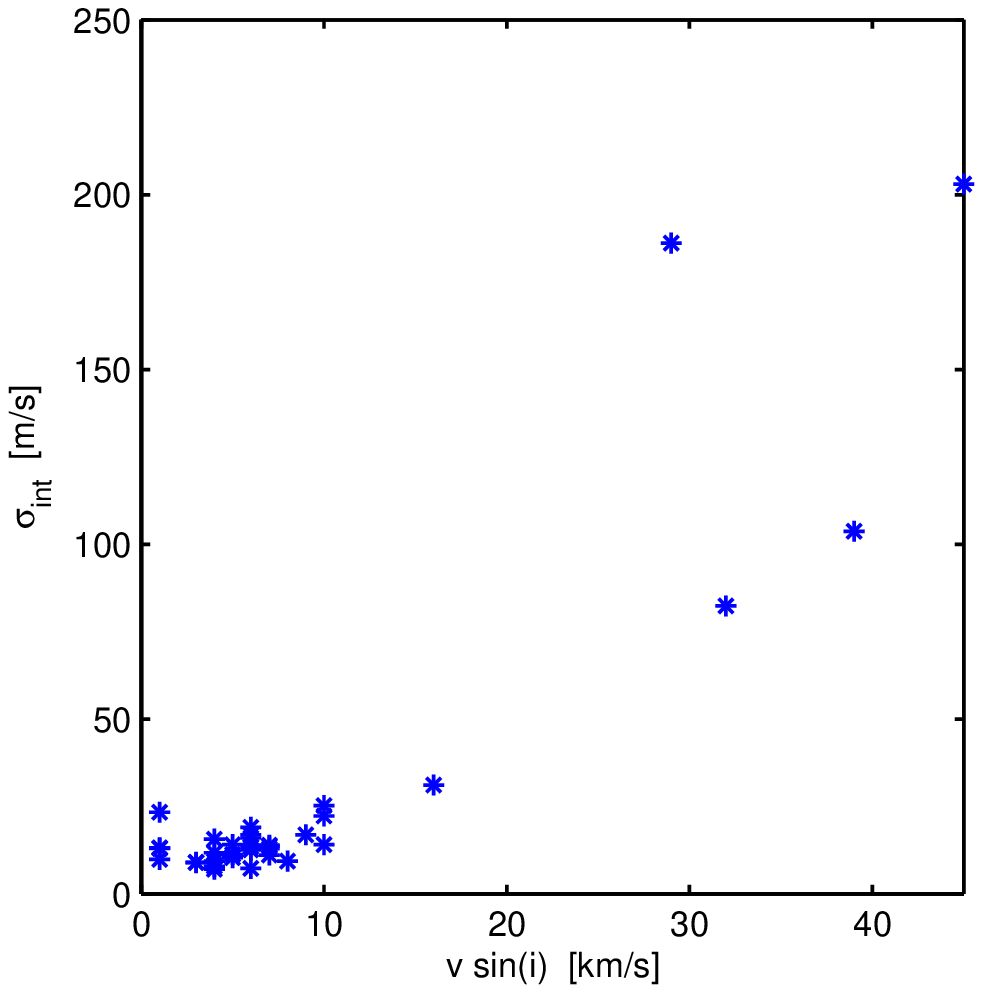}{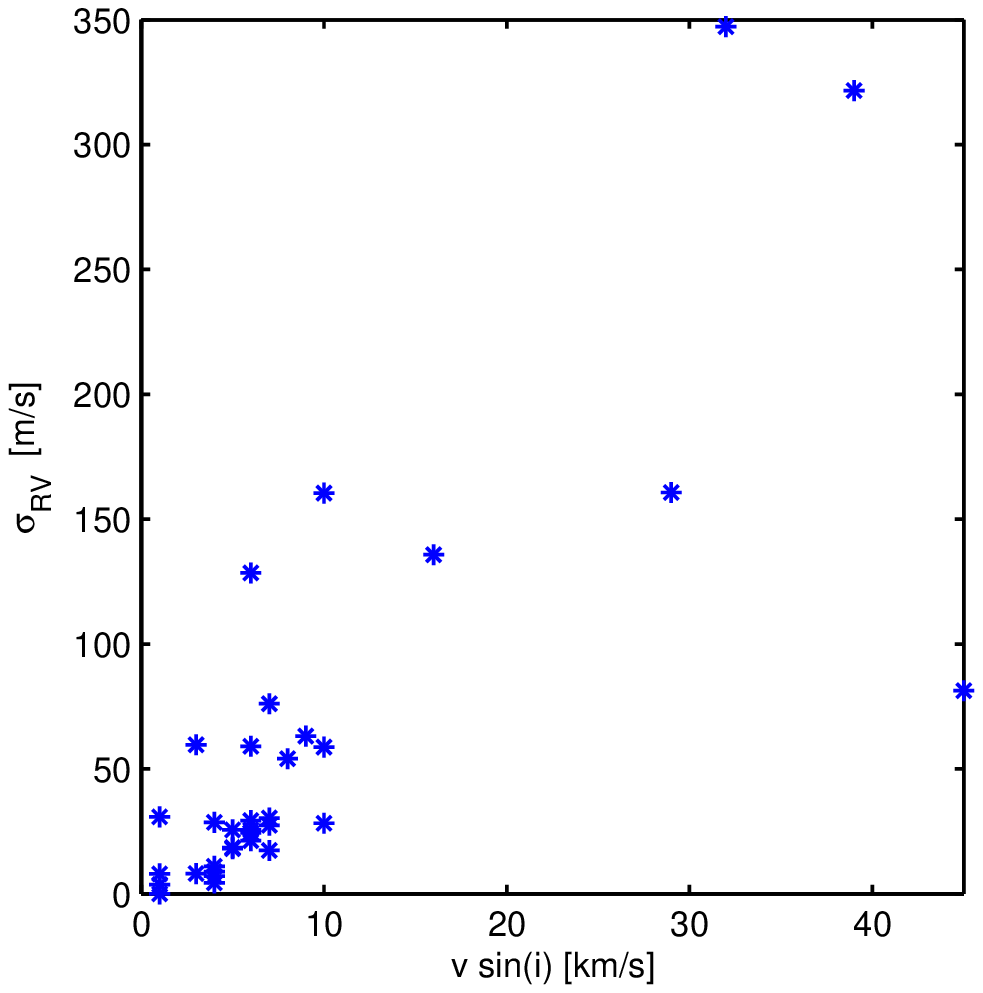}
\caption{(left) Plot of the average internal errors $\sigma_{int}$ 
as a function of  the stellar rotational velocity $v\sin i$. 
(right) $\sigma_{RV}$ vs $v\sin i$; see the text for the definition 
of $\sigma_{RV}$. In this analysis only 34 stars have been 
considered excluding stars with less than 20 data points and stars 
which  show clear RV-trend.  } 
\end{center}
\end{figure}

\subsection{RV-jitter}

Apart from the fact that the RV's for rapidly rotating stars are more 
difficult to determine, these stars are also more active. In order to 
demonstrate this, we consider the $\sigma_{RV}$ which is defined 
as the square-root of the difference between the standard
deviation of the observed RV's ($\sigma_{obs}$) and the   
$\sigma_{int}$ squared. $\sigma_{RV}$ thus is a measure of 
the RV-variations presumably due to stellar activity. Fig. 2 (right) 
shows $\sigma_{RV}$ against $v\sin i$. It can clearly be seen 
that stars of larger  $v\sin i$ also show larger $\sigma_{RV}$-values.
 On the other hand the majority of  stars in our sample which have  
$v\sin i \leq 10 $ km/s show a $\sigma_{RV}$ lower than 35 m/s. 
Such a level of `noise' in the RV-measurements, even if larger than 
typical values for old dwarf stars of the same spectral type 
(Santos et al. 2000), still allows the detection of a RV-signal of 
most of the known exo-planets. For instance,  a planet with 
$M \sin i$ = 1 $M_J$  in a circular orbit around  a 1 $M_{\sun}$ 
star with a period $P=10(100)$ days induces a stellar wobble with 
a RV semi-amplitude of $K=94(44)$ m/s which is larger than the 
scatter caused by activity for a star with a  $v\sin i \leq 10 $ km/s. 

\begin{figure}[th]
\plotone{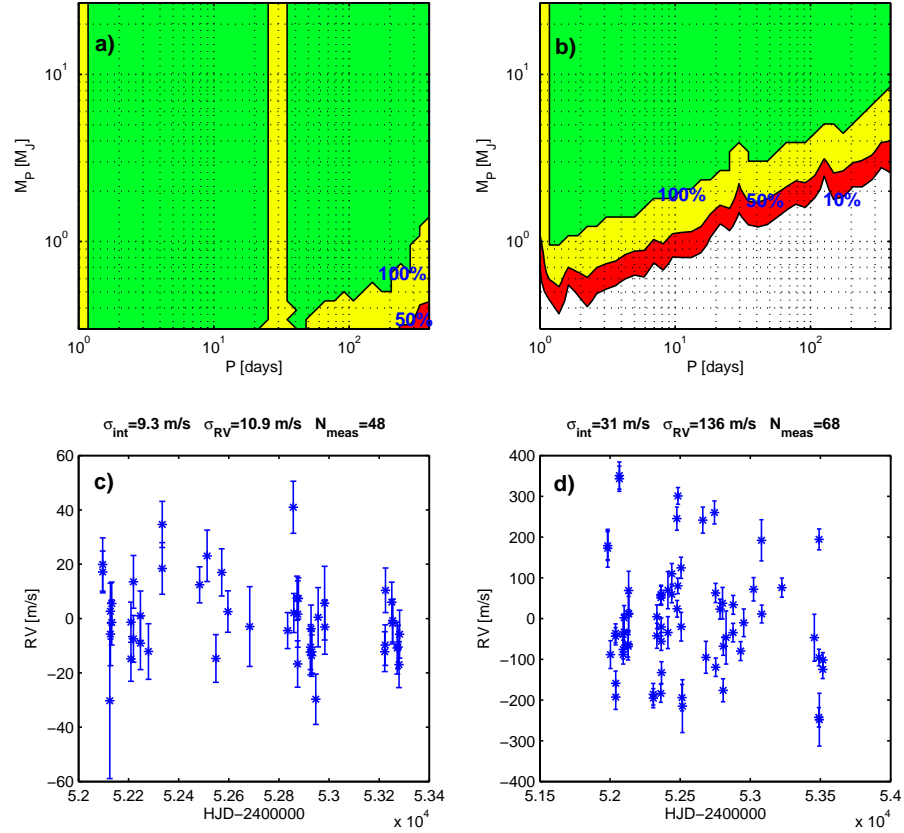}
\caption{(lower panels) RV vs HJD (heliocentric Julian date) 
for two stars showing a low (c) and high (d) level of jitter 
respectively. (upper panels) The detection limits in the 
mass-period diagram. Iso-probability curve for the 
100\%, 50\% and 10\% levels are plotted. In the green 
zone we are virtually sure that a planet would have 
been detected.} 
\end{figure}

\subsection{Detection limits}

In order to derive upper limits for the detection of planets 
in our sample, we carried out simulations. For each point on 
a grid of values of the orbital period $P$ and the planetary 
mass $M\sin i$, assuming for the star a mass on the basis of 
its spectral type, we generate a sinusoidal RV curve. The curve 
is then sampled in the same way as the real data for that
 star, obtaining a set of simulated RV values.In the next step, we
 add noise of normal distribution to the  data. The noise is scaled 
so that the  $\sigma$ of the simulated data is equal to the 
observed $\sigma_{obs}$. The whole procedure is repeated 
5000 times, varying the phases randomly. For each of the 5000
 simulated data-sets we then derive the value of the periodogram 
($S$) for that period $P$ (Scargle 1982) and thus obtain for a 
given $P$ and $M \sin i$ a distribution of $S$, which is 
compared to a similar distribution for $M \sin i = 0$. If the two 
distributions do not overlap, we conclude that a planet of that 
$P$ and $M \sin i$ can be excluded. Similarly, from the overlap 
of the two distributions, we can calculate the probability for 
excluding such planet.

The results of our simulations for two opposite cases are shown 
in Fig.3. Quite remarkably in the best conditions (panel $a$) we 
easily would have been able to detect a planet with 
$M\sin i = 1M_{J}$  for a period as large as 300 days. Besides,
 even in the worst case (panel $b$) thanks to the high number 
of data points ($N_{meas}=68$) we are still sensible to very 
hot Jupiter-mass planets. 

\subsection{A young exo-planet candidate}

\begin{figure}[th]
\plotone{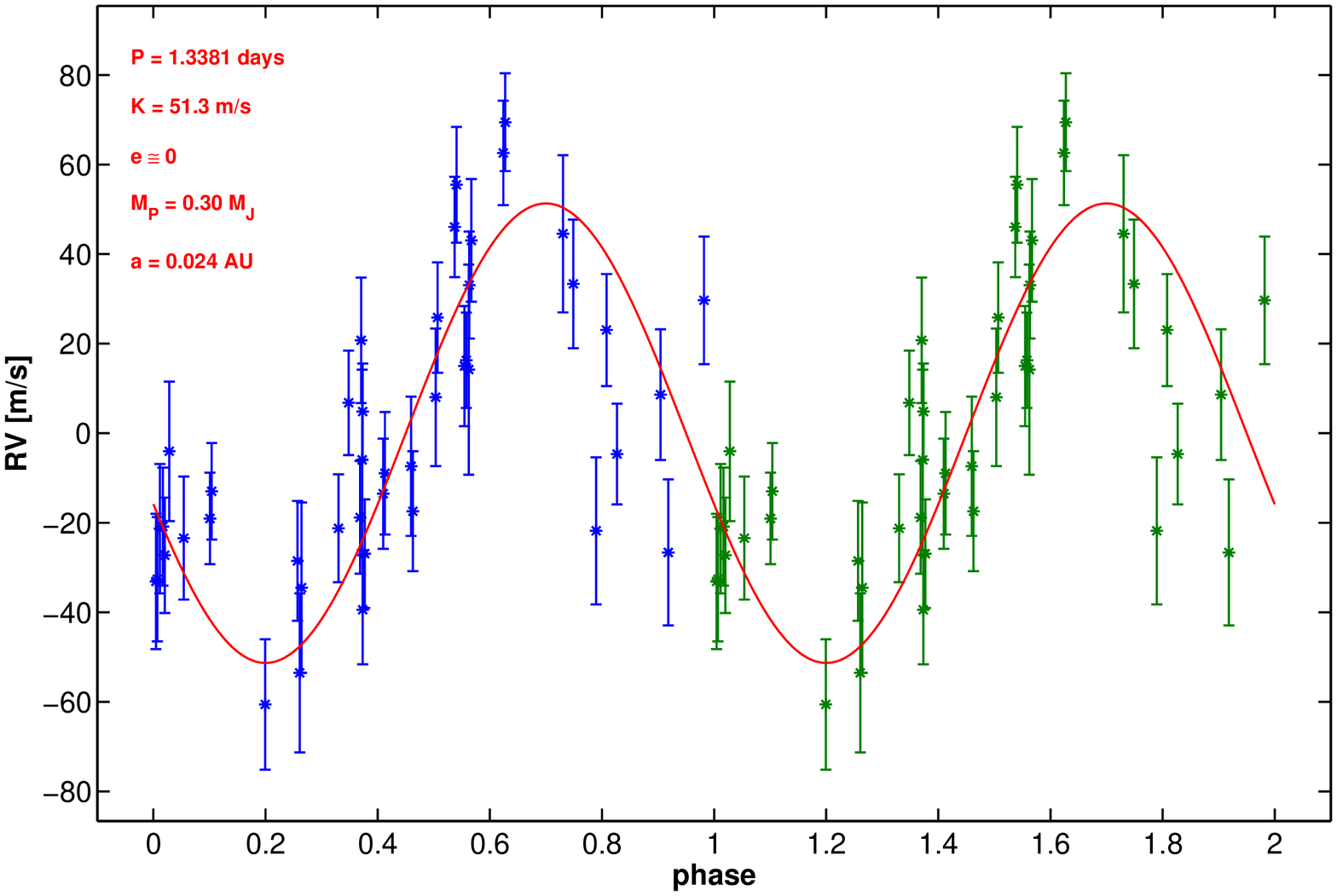}
\caption{The RV measurements for a G0 star in our sample 
phase-folded with a P=1.3381 days period. 
The best-fit RV curve and corresponding orbital parameters are 
reported. } 
\end{figure}

In the whole sample we have found only three objects which 
show a long period RV trend compatible with the presence of a 
stellar companion and one  young planet candidate. The RV 
data set for all these stars has been analysed in the same 
way: we calculated the Scargle periodogram and in 
correspondence to the periods with higher peaks we estimate 
the weighted least squares best-fit orbital solution. 
The phase-folded RV curve for our candidate planet is shown 
in Fig.4 together with the best-fit orbital parameters. As we 
know that RV variations can also be induced by stellar 
activity, we analysed the Hipparcos photometry and found 
a P=1.6477 days periodicity. It is hence possible that  the RV 
as well as the photometric variations originate from a $\sim$ 1.5 
days stellar rotational period. However the star does not 
present enhanced X-ray luminosity and strong emission in the 
cores of the CaII H and K lines, which are typical of fast 
rotators, and the $v\sin i = 4$km/s  would imply that we are 
observing the star almost pole-on. Further simultaneous 
spectroscopic and photometric observations will possibly 
confirm or discard the planet hypothesis.

\acknowledgments{
M. Esposito's work was performed under the auspices 
of the EU, which has provided financial support to the 
``Dottorato di Ricerca Internazionale
 in Fisica della Gravitazione ed Astrofisica'' of the Salerno 
University, through the ``Fondo Sociale Europeo, Misura III.4''.
}

\end{document}